\documentclass[reprint,prl,amsmath,amssymb,superscriptaddress,noffotinbib,preprintnumbers]{revtex4-1}
\usepackage{url}
\usepackage{color}
\usepackage{isotope}
\usepackage{pstricks}
\usepackage{colordvi}
\usepackage{epsfig}
\usepackage{graphicx}
\usepackage{dcolumn}
\usepackage{bm}
\usepackage{amsmath}
\usepackage{amssymb}
\usepackage{hyperref}
\usepackage{slashed}
\usepackage{lineno}
\usepackage{isotope}

\hypersetup{
    colorlinks=true, 
    linkcolor=blue,  
   citecolor=blue
}

\begin{document}


\title{First Measurement of the {Ti}$(e,e^\prime){\rm X}$ Cross Section at Jefferson Lab}

\author{H.~Dai} \affiliation{Center for Neutrino Physics, Virginia Tech, Blacksburg, Virginia 24061, USA}
\author{M.~Murphy} \affiliation{Center for Neutrino Physics, Virginia Tech, Blacksburg, Virginia 24061, USA} 
\author{V.~Pandey} \email{vishvas.pandey@vt.edu} \affiliation{Center for Neutrino Physics, Virginia Tech, Blacksburg, Virginia 24061, USA} 
\author{D.~Abrams} \affiliation{Department of Physics, University of Virginia, Charlottesville, VA, 22904, USA}
\author{D.~Nguyen} \affiliation{Department of Physics, University of Virginia, Charlottesville, VA, 22904, USA}
\author{B.~Aljawrneh} \affiliation{North Carolina Agricultural and Technical State University, Greensboro, North Carolina, 27401, USA}
\author{S.~Alsalmi} \affiliation{Kent State University, Kent, Ohio, 44242, USA}
\author{A.~M.~Ankowski} \altaffiliation{Present address: SLAC National Accelerator Laboratory, Stanford University, Menlo Park, CA, 94025, USA} \affiliation{Center for Neutrino Physics, Virginia Tech, Blacksburg, Virginia 24061, USA} \affiliation{SLAC National Accelerator Laboratory, Stanford University, Menlo Park, CA, 94025, USA} 
\author{J.~Bane} \affiliation{The University of Tennessee, Knoxville, Tennessee 37996, USA}
\author{S.~Barcus} \affiliation{The College of William and Mary, Williamsburg, Virginia, 23187, USA}
\author{O.~Benhar} \affiliation{INFN and Dipartimento di Fisica, Sapienza Universit\`{a} di Roma, I-00185 Roma, Italy}
\author{V.~Bellini} \affiliation{INFN, Sezione di Catania, Catania, 95123, Italy}
\author{J.~Bericic} \affiliation{Thomas Jefferson National Accelerator Facility, Newport News, Virginia 23606, USA}
\author{D.~Biswas} \affiliation{Hampton University, Hampton, Virginia, 23669, USA}
\author{A.~Camsonne} \affiliation{Thomas Jefferson National Accelerator Facility, Newport News, Virginia 23606, USA}
\author{J.~Castellanos} \affiliation{Florida International University, Miami, FL, 33181, USA}
\author{J.-P.~Chen} \affiliation{Thomas Jefferson National Accelerator Facility, Newport News, Virginia 23606, USA}
\author{M.~E.~Christy} \affiliation{Hampton University, Hampton, Virginia, 23669, USA}
\author{K.~Craycraft} \affiliation{The University of Tennessee, Knoxville, Tennessee 37996, USA}
\author{R.~Cruz-Torres} \affiliation{Massachusetts Institute of Technology, Cambridge, Massachusetts, 02139, USA}
\author{D.~Day} \affiliation{Department of Physics, University of Virginia, Charlottesville, VA, 22904, USA}
\author{S.-C.~Dusa} \affiliation{Thomas Jefferson National Accelerator Facility, Newport News, Virginia 23606, USA}
\author{E.~Fuchey} \affiliation{University of Connecticut, Storrs, Connecticut 06269, USA}
\author{T.~Gautam} \affiliation{Hampton University, Hampton, Virginia, 23669, USA}
\author{C.~Giusti} \affiliation{Dipartimento di Fisica, Universit\`{a} degli Studi di Pavia and\\
INFN, Sezione di Pavia,  I-27100 Pavia, Italy}
\author{J.~Gomez} \affiliation{Thomas Jefferson National Accelerator Facility, Newport News, Virginia 23606, USA}
\author{C.~Gu} \affiliation{Department of Physics, University of Virginia, Charlottesville, VA, 22904, USA}
\author{T.~Hague} \affiliation{Kent State University, Kent, Ohio, 44242, USA}
\author{J.-O.~Hansen} \affiliation{Thomas Jefferson National Accelerator Facility, Newport News, Virginia 23606, USA}
\author{F.~Hauenstein} \affiliation{Old Dominion University, Norfolk, VA 23529, USA}
\author{D.~W.~Higinbotham} \affiliation{Thomas Jefferson National Accelerator Facility, Newport News, Virginia 23606, USA}
\author{C.~Hyde} \affiliation{Old Dominion University, Norfolk, VA 23529, USA}
\author{C.~M.~Jen} \affiliation{Center for Neutrino Physics, Virginia Tech, Blacksburg, Virginia 24061, USA}
\author{C.~Keppel} \affiliation{Thomas Jefferson National Accelerator Facility, Newport News, Virginia 23606, USA}
\author{S.~Li} \affiliation{University of New Hampshire, Durham, New Hampshire, 03824, USA}
\author{R.~Lindgren} \affiliation{University of Virginia, Charlottesville, Virginia, 22908, USA}
\author{H.~Liu} \affiliation{Columbia University, New York, New York, 10027, USA}
\author{C.~Mariani} \affiliation{Center for Neutrino Physics, Virginia Tech, Blacksburg, Virginia 24061, USA} 
\author{R.~E.~McClellan} \affiliation{Thomas Jefferson National Accelerator Facility, Newport News, Virginia 23606, USA} 
\author{D.~Meekins} \affiliation{Thomas Jefferson National Accelerator Facility, Newport News, Virginia 23606, USA}
\author{R.~Michaels} \affiliation{Thomas Jefferson National Accelerator Facility, Newport News, Virginia 23606, USA}
\author{M.~Mihovilovic} \affiliation{Jozef Stefan Institute, Ljubljana 1000, Slovenia}
\author{M.~Nycz} \affiliation{Kent State University, Kent, Ohio, 44242, USA}
\author{L.~Ou} \affiliation{Massachusetts Institute of Technology, Cambridge, Massachusetts, 02139, USA}
\author{B.~Pandey} \affiliation{Hampton University, Hampton, Virginia, 23669, USA}
\author{K.~Park} \affiliation{Thomas Jefferson National Accelerator Facility, Newport News, Virginia 23606, USA}
\author{G.~Perera} \affiliation{University of Virginia, Charlottesville, Virginia, 22908, USA}
\author{A.J.R.~Puckett} \affiliation{University of Connecticut, Storrs, Connecticut 06269, USA}
\author{S.~\v{S}irca} \affiliation{University of Ljubljana, Ljubljana, 1000, Slovenia} \affiliation{Jozef Stefan Institute, Ljubljana 1000, Slovenia}
\author{T.~Su} \affiliation{Kent State University, Kent, Ohio, 44242, USA}
\author{L.~Tang} \affiliation{Hampton University, Hampton, Virginia, 23669, USA}
\author{Y.~Tian} \affiliation{Shandong University, Shandong, 250000, China}
\author{N.~Ton} \affiliation{University of Virginia, Charlottesville, Virginia, 22908, USA}
\author{B.~Wojtsekhowski} \affiliation{Thomas Jefferson National Accelerator Facility, Newport News, Virginia 23606, USA}
\author{S.~Wood} \affiliation{Thomas Jefferson National Accelerator Facility, Newport News, Virginia 23606, USA}
\author{Z.~Ye} \affiliation{Physics Division, Argonne National Laboratory, Argonne, Illinois 60439, USA}
\author{J.~Zhang} \affiliation{University of Virginia, Charlottesville, Virginia, 22908, USA}

\collaboration{The Jefferson Lab Hall A Collaboration}


\begin{abstract}
To probe CP violation in the leptonic sector using GeV energy neutrino beams in current and future experiments using argon detectors, precise models of the complex underlying neutrino and antineutrino interactions are needed. The E12-14-012 experiment at Jefferson Lab Hall A was designed to perform a combined analysis of inclusive and exclusive electron scatterings on both argon ($N = 22$) and titanium ($Z = 22$) nuclei using GeV energy electron beams. The measurement on titanium nucleus provides essential information to understand the neutrino scattering on argon, large contribution to which comes from scattering off neutrons.
Here we report the first experimental study of electron-titanium scattering as double differential cross section at beam energy $E=2.222$~GeV and electron scattering angle $\theta = 15.541$~deg, measured over a broad range of energy transfer, spanning the kinematical regions in which quasielastic scattering and delta production are the dominant reaction mechanisms. The data provide valuable new information needed to develop accurate theoretical models of the electromagnetic and weak cross sections of these complex nuclei in the kinematic regime of interest to neutrino experiments.
\end{abstract}
\preprint{JLAB-PHY-18-2656}
\preprint{SLAC-PUB-17200}
%
%
\maketitle
\par The interpretation of the data collected by experimental studies of neutrino oscillations demands a fully quantitative description of neutrino interactions with the atomic nuclei comprising the detector~\cite{Benhar:2015wva}. Current and future neutrino experiments, such as the short- (SBN)~\cite{SBNProposal:2015} and the long-baseline (DUNE)~\cite{DUNEReport:2015} neutrino programs, will use detectors based on the liquid-argon time-projection chambers (LArTPCs) technology. In order to achieve the precision goals of these programs, the treatment of nuclear effects, which has been recognized as a major source of systematic uncertainty in ongoing experiments~\cite{Abe:2013}, has to be addressed. Realistic models of both neutrino- and antineutrino-argon interactions will be even more critical to future experiments, such as DUNE, aimed at pinning down charge-parity (CP) symmetry violation in the leptonic sector, because its determination with few percent precision requires accurate measurements of both neutrino and antineutrino oscillations. Failing to achieve this goal in a timely manner will deeply affect the sensitivity of DUNE to measure the CP violating phase ($\delta_{CP}$), as discussed in~\cite{Ankowski:2015jya}. 
\par Since the description of nuclear effects in a non isospin-symmetric nucleus, such as  argon, must take into account the differences in the shell-model states occupied by protons and neutrons, and the information on neutrons cannot be directly extracted, one has to resort to studies of titanium nucleus. Owing to the fact that the neutron spectrum of \isotope[40][18]{Ar} is mirrored by the proton spectrum of Ti, the Ti data will give access to the neutron spectral function of argon. Given the scarcity of electron-argon scattering experiments\textemdash the only available data on argon being the inclusive spectrum measured at Frascati National Laboratory using the electron-positron collider ADONE and a jet target~\cite{Anghinolfi:1995}\textemdash we performed a dedicated experiment at Jefferson Lab (JLab), aimed at measuring inclusive and exclusive cross-sections on both argon and titanium targets, and extracting the proton [via \isotope[][]{Ar}$(e, e^\prime p)$] and neutron [via \isotope[][]{Ti}$(e, e^\prime p)$)] spectral functions of the argon nucleus in the kinematical region in which shell-model dynamics is dominant~\cite{Proposal:2014}.
\par Electron-scattering experiments have provided a wealth of information on the nuclear response to electromagnetic interactions over a broad kinematic regime. Static form factors and charge distributions have been extracted from elastic scattering data, while measurements of inelastic cross sections have allowed for systematic studies of the dynamic response functions, which shed light on the role played by different reaction mechanisms. Finally, with the advent of continuous beam accelerators, a number of exclusive processes have been analyzed to unprecedented precision. The availability of the body of electron-nucleus scattering data has been essential to the development of theoretical models. Most notably, experimental studies of the  $(e, e^\prime p)$ process\textemdash in which the scattered electron and the knocked-out proton are detected in coincidence\textemdash have provided detailed information on proton spectral functions~\cite{benhar:NPN,Frullani,Dieperink,Boffi}, the knowledge of which is needed to obtain the nuclear cross sections within the factorization scheme underlying the impulse approximation (IA). The description based on the IA and the spectral function formalism~\cite{Benhar:2005} has been successful in describing inclusive electron-scattering data in a variety of kinematic regimes~\cite{Benhar:Review-2008,Ankowski:2015,Rocco:2016}, and has recently been be extended to the analysis of neutrino scattering~\cite{NPA,Coletti,Veneziano,Vagnoni:2017}. However, due to the scarcity of available inclusive cross sections for argon and titanium, theoretical models of nuclear effects in electroweak interactions~\cite{Lovato:2016, Meucci:2014, Lalakulich:2012, Nieves:2012, Martini:2010, Pandey:2014, Megias:2016} cannot be currently validated in the kinematical region relevant for neutrino experiments.
\par The experiment E12-14-012 planned to perform a combined analysis of Ar and Ti inclusive and exclusive reactions collected high statistics data in JLab Hall A during February-March 2017. Here, we report the first results of the experiment, including the {Ti}$(e,e^\prime){X}$ cross section at beam energy $E = 2.222$~GeV and electron scattering angle $\theta = 15.541$~deg and the corresponding electron-carbon cross section. Note that this is the first double differential {Ti}$(e,e^\prime){X}$ cross section measured at the kinematics relevant for neutrino experiments, the previous studies on titanium target include~\cite{Heisenberg:1971, Selig:1988, Romberg:1971}. The measurement of the  {C}$(e,e^\prime){X}$ cross section allowed a comparison with previous experiments, as well as a careful study of systematic uncertainties. 
In the $(e,e^\prime)$ process $ e + A \rightarrow e^\prime  + X $,
an electron of four-momentum $k \equiv (E, {\bf k})$ scatters off a nuclear target $A$. The energy and emission angle of the scattered electron of four-momentum $k^\prime\equiv (E^\prime, {\bf k}^\prime )$ are measured, while the hadronic final state is left undetected. The squared four-momentum transfer in the process is $q^2=-Q^2$, with $q = k - k^\prime \equiv (\omega, {\bf q})$.
\par The electron beam was provided by the Continuous Electron Beam Accelerator Facility (CEBAF) at JLab, with currents in excess of 10~$\mu$A. The beam current was monitored using two Beam Current Monitors (BCMs), which are resonant radio-frequency cavities. The position of the beam was monitored by two similar cavity types, Beam Position Monitors (BPMs). 
Beam size was measured with harp scanners, which also allowed the cavity monitor calibration. Beam position determination is important for vertex reconstruction and momentum calculation of the scattered electron. The beam was rastered, with a 2~mm $\times$ 2~mm raster system, in both vertical and horizontal directions, to reduce the beam current density and hence eliminate the possibility of melting the solid foil targets and minimize the local heating of the cryogenic hydrogen target. Both the carbon and titanium targets were foils of natural isotope composition, with a thickness of 0.167$\pm$0.001~g/cm$^2$ and 0.729$\pm$0.001~g/cm$^2$, respectively. 
\par The scattered particles were momentum analyzed by two nearly identical spectrometers\textemdash the Left and a Right High-Resolution Spectrometers (HRSs)\textemdash equipped with detectors for tracking, timing and particle ID. The HRSs consist of 4 magnets (3 superconducting and 1 resistive) in a QQDQ configuration, where the Q indicates a quadrupole magnet and the D indicates a dipole magnet. This arrangement provided a large acceptance for both the angle and momentum, with a relative momentum resolution of $\sim$10$^{-4}$, and pointing accuracy and angular resolution of $\sim$10$^{-4}$~m and $\sim$10~mrad, respectively. The detector package, slightly updated with respect to the one in Ref.~\cite{Alcorn:2004}, consisting of vertical drift chambers (VDCs), threshold \v{C}erenkov counters, scintillator detectors and a lead-glass calorimeters, provides data-acquisition triggering, tracking, and particle identification.
\par The scattered electrons were detected in the Left HRS positioned at $\theta = 15.541$~deg. The data acquisition was triggered, with an efficiency of 99.9\%, when an electron fired the two scintillator detectors planes (with a logical {\textsc{and}}) simultaneously with a signal in the gas \v{C}erenkov detector. The electrons were identified by a gas threshold \v{C}erenkov detector, mounted between two scintillator detector planes, with 99.9\% efficiency and negligible pion contamination. The track trajectories were reconstructed in the detector stack using the VDCs with efficiencies of $\sim$ 95\% for C and $\sim$ 92\% for Ti, and then transported to the target utilizing a fitted reconstruction matrix obtained from a special optics calibration run. We required only one VDC track reconstructed in the final state for simplicity and purity purposes. The number of VDC reconstructed tracks in the case of the Ti target is slightly higher than for the C target, the difference in respective VDC efficiencies was as expected. 
\par The cross section is extracted first computing the yield defined in both data and simulation as:
\begin{equation}
\text{Yield}^i=(N_S^i \times {\text{DAQ}}_{\text{pre-scale}}) / (N_e \times LT \times \epsilon).
\end{equation}
Here, $i$ is the $i$th bin in $E^\prime$, $N_S^i$ represents the number of scattered electrons, $N_e$ is the total number of electrons on the target, $LT$ is the live-time, $\epsilon$ is the total efficiency and $\text{DAQ}_{\text{pre-scale}}$ is a factor that determines what fraction of the events gets recorded. The cross section in each bin $i$ is computed as the product of the Monte Carlo (MC) cross section~\cite{Arrington1} times the ratio of the data to simulation yields. The MC cross section is a fit to existing data, including preliminary Hall C~\cite{HallC} data and includes radiative corrections computed using the peaking approximation~\cite{MoTSAI} and Coulomb corrections implemented with an effective momentum approximation~\cite{Aste}.
\begin{figure}[t!]
\centering
\includegraphics[width=0.85\columnwidth]{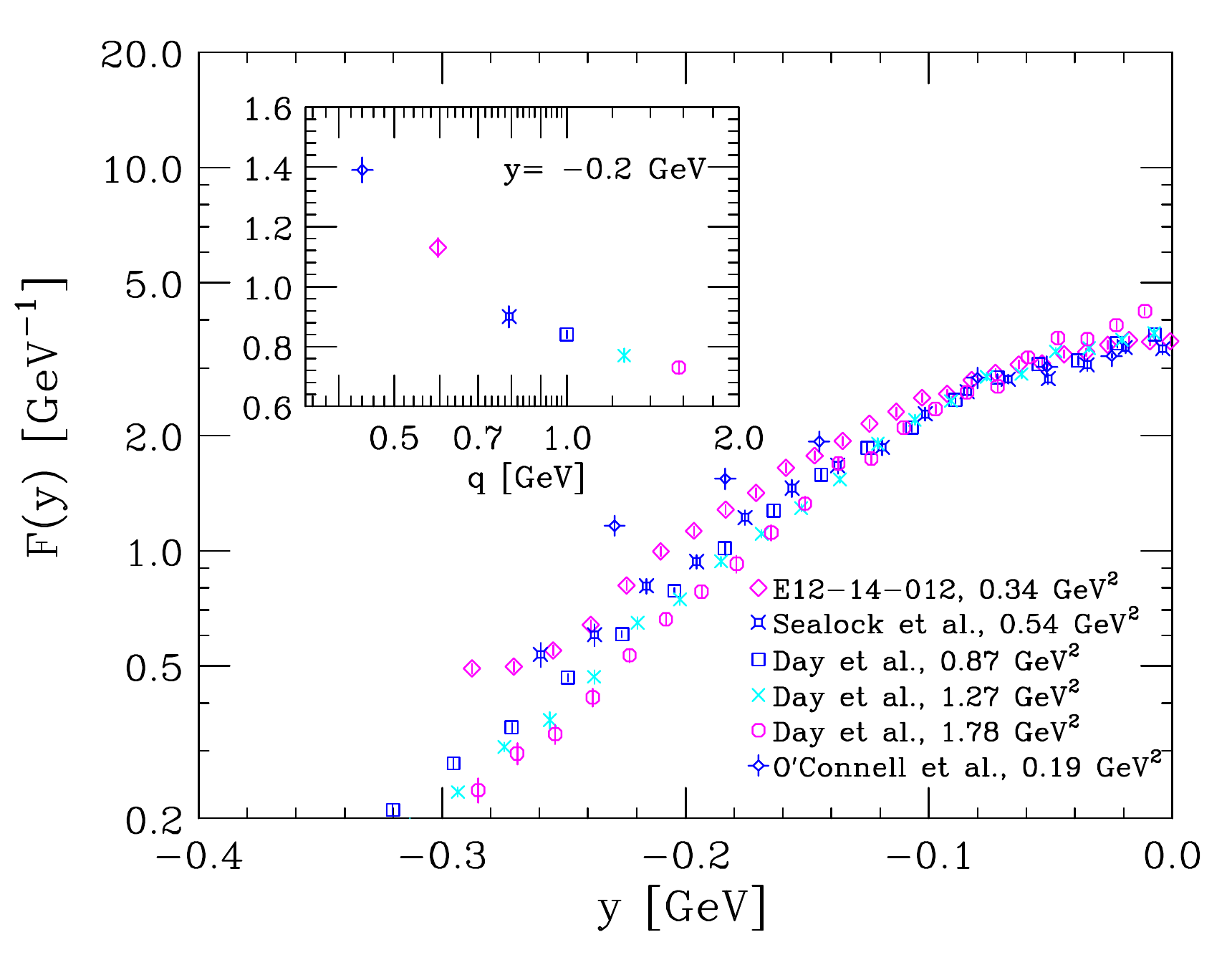}
\caption{(color online). Comparison between the scaling function $F(y)$ obtained from the E12-14-012 data on C, represented by diamonds, and those obtained from the data of O'Connell {\it et al.}~\cite{oconnell}, Sealock {\it et al.}~\cite{sealock}, and Day {\it et al.}~\cite{day}. The inset shows the momentum transfer dependence of $F(y)$ at fixed $y=-0.2$ GeV. The data sets are labeled by the value of $Q^2$ corresponding to the top of the quasielastic peak.}
\label{scaling_function}
\end{figure}
\par Over nearly five decades, a number of measurements of the electron-carbon cross section have been performed at different electron-scattering facilities around the world. A compilation of the available inclusive data can be found in Ref.~\cite{Day:Archive}. In order to put our results in perspective, in Fig.~\ref{scaling_function} we compare the $y$-scaling function~\cite{yscaling},  $F(y)$, obtained from the cross section measured by the E12-14-012 experiment to those obtained from the data of Refs.~\cite{oconnell,sealock,day}, spanning a kinematical range corresponding to $0.20 \lesssim Q^2 \lesssim 1.8$~GeV$^2$. The occurrence of scaling in the variable, i.e. the observation that $F(y)$ becomes independent of the momentum transfer $|{\bf q}|$ in the limit of large $|{\bf q}|$, indicates that quasielastic scattering is the dominant reaction mechanism and final state interactions (FSI) between the knocked out nucleon and the residual nucleus are negligible. The scaling variable $y$, whose definition is given in Ref.~\cite{yscaling}, can be loosely identified with the component of the initial nucleon momentum parallel to the momentum transfer.
\par The main panel of Fig.~\ref{scaling_function} clearly shows that the data exhibit a remarkable scaling behavior at $y\approx0$, corresponding to $\omega \approx Q^2/2M$, where $M$ is the nucleon mass, while sizable scaling violations, to be mainly ascribed to FSI, are observed at large negative values of $y$. The momentum transfer dependence of $F(y)$ at $y=-0.2$ GeV, illustrated in the inset, demonstrates that in the kinematical setup of our experiment, corresponding to $|{\bf q}|\approx 600$~MeV, the effects of FSI are still significant. Overall, Fig.~\ref{scaling_function} shows that our results are fully consistent with those of previous experiments.
\begin{figure}[t!]
\centering
\includegraphics[width=0.8\columnwidth]{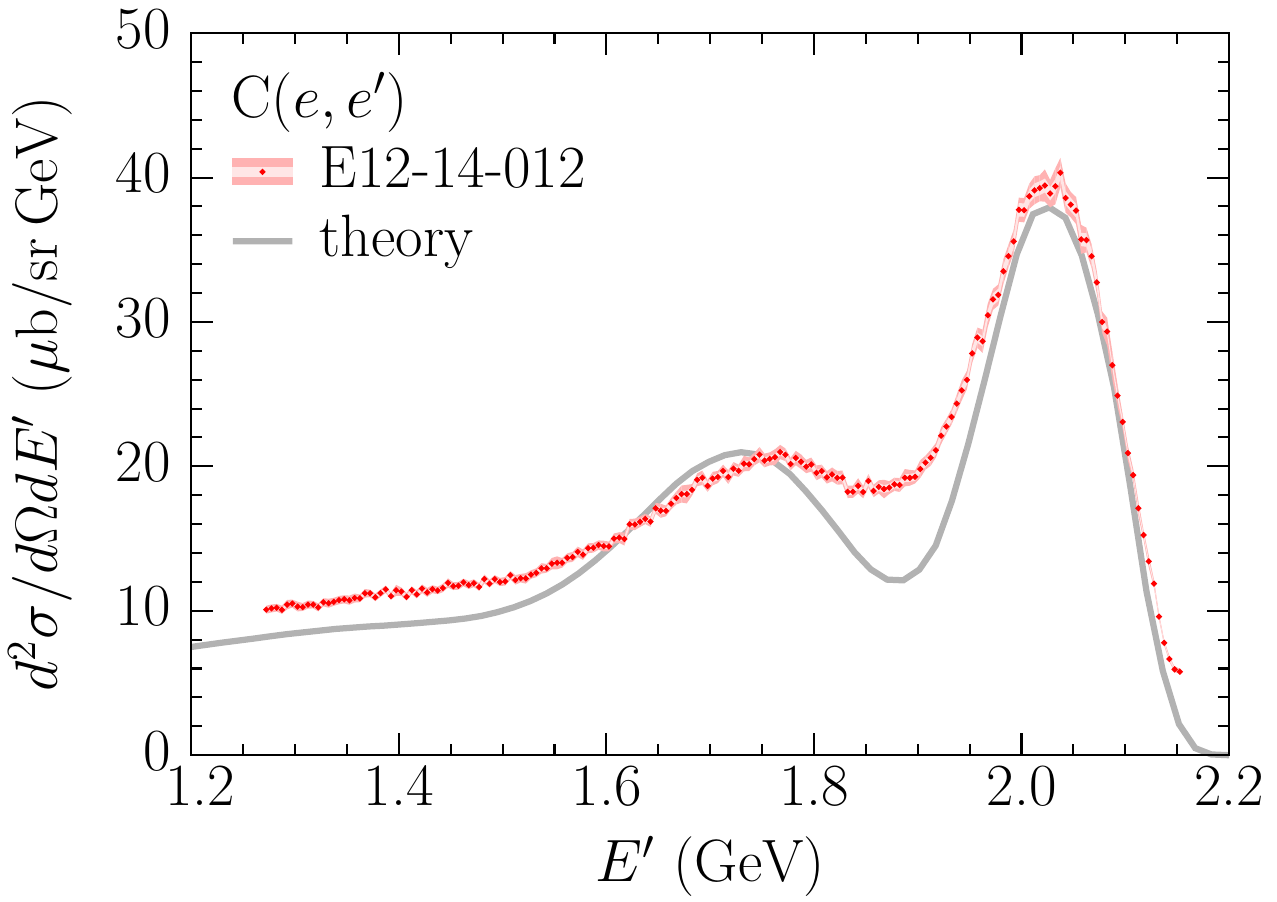}
\caption{(color online). Double differential cross section for the {C}$(e,e^\prime)$ process measured at beam energy of 2.222~GeV and scattering angle of 15.541~deg. The inner and outer uncertainty bars correspond to statistical and total uncertainties, respectively. The solid line represents the theoretical results obtained within the formalism described in Refs.~\cite{Benhar:2005,Ankowski:2015,Rocco:2016,gosix}.}
\label{xsec_C}
\end{figure}
\begin{table}[h!]
\caption{Contributions to the uncertainties associated with the measured C$(e,e^\prime )$ cross sections. Numbers represents upper limit or range for the uncertainties that vary between kinematical regions.}
\label{tab:syst}
\begin{center}
\begin{tabular}{l c c c}
\\
\hline\hline
{1. Total statistical uncertainty}Ê &						& 1.2\% &\\
{2. Total systematic uncertainty}Ê &						& 2.0--2.9\% &\\
\phantom{1. }a.~Beam charge \& Beam Energy & 			& 0.3\% &\\
\phantom{1. }b.~Beam offset $x$\&$y$ & 					& 0.1\%--0.4\% &\\
\phantom{1. }c.~Target thickness & 						& 0.1\%--0.4\% &\\
\phantom{1. }d.~HRS offset $x$\&$y$Ê$+$ Optics & 			& 1.3\%--2.0\% &\\
\phantom{1. }e.~Acceptance cut($\theta$,$\phi$,$dp/p$) & 	& 1.0\%--1.4\% &\\
\phantom{1. }f.~Calorimeter \& \v{C}erenkov cuts & 		        & 0.01\%--0.02\% &\\
\phantom{1. }g.~Cross Section Model &					& 0.1\%--0.2\%&\\
\phantom{1. }h.~Radiative $+$Coulomb Corr.Ê & 		        & 1.0--1.3\% &\\
\hline
\hline\\[-20pt]
\end{tabular}
\end{center}
\end{table}
\par Figure~\ref{xsec_C} shows the measured C$(e,e^\prime )$ cross section as a function of the energy of the scattered electron, ranging from $\sim$1.2~GeV to $\sim$2.2~GeV with error bars up to $\sim$2.5\%, corresponding to the statistical (1.2\%) and systematic (2.2\%) uncertainties summed in quadrature. It can be seen that the kinematical coverage includes both the quasielastic and delta-production peaks, and extends to the region in which the contribution of deep-inelastic scattering becomes appreciable. The statistical uncertainty includes beam charge (0.03\%), detector and trigger efficiencies (0.1\%), DAQ live-time (0.02\%), VDC, and VDC track reconstruction efficiencies (0.1\%) and uncertainties due to the charge-symmetric background prediction~\cite{Tvaskis} (0.01\%). A detailed list of the systematic uncertainties is given in Table~\ref{tab:syst}. All uncertainties are considered as fully uncorrelated. This new high precision C$(e,e^\prime )$ data not only allowed us to carefully test our analysis framework and study systematics but also provides a vital information for the neutrino experiments that use carbon targets such as the long-baseline neutrino experiment T2K~\cite{Abe:2011}, NOvA~\cite{NOvA:2017} and neutrino interaction experiment MINERvA~\cite{MINERvA:2017}.

The solid line of Fig.~\ref{xsec_C} represents theoretical results obtained within the scheme described in Refs.~\cite{Benhar:2005,Ankowski:2015,Rocco:2016,gosix}, based on the factorization {\em ansatz} dictated by the IA and the spectral function formalism. Note that this approach does not involve any adjustable parameters, and allows for a consistent inclusion of single-nucleon interactions\textemdash both elastic and inelastic\textemdash and meson-exchange current (MEC) contributions. The effects of FSI on the quasielastic cross section has been taken into account following the procedure developed in  Ref.~\cite{gosix}. A detailed account of the calculation of the electron-carbon cross section will be provided in a forthcoming paper~\cite{Benhar_Coppini}.
\begin{figure}[t!]
\centering
\includegraphics[width=0.8\columnwidth]{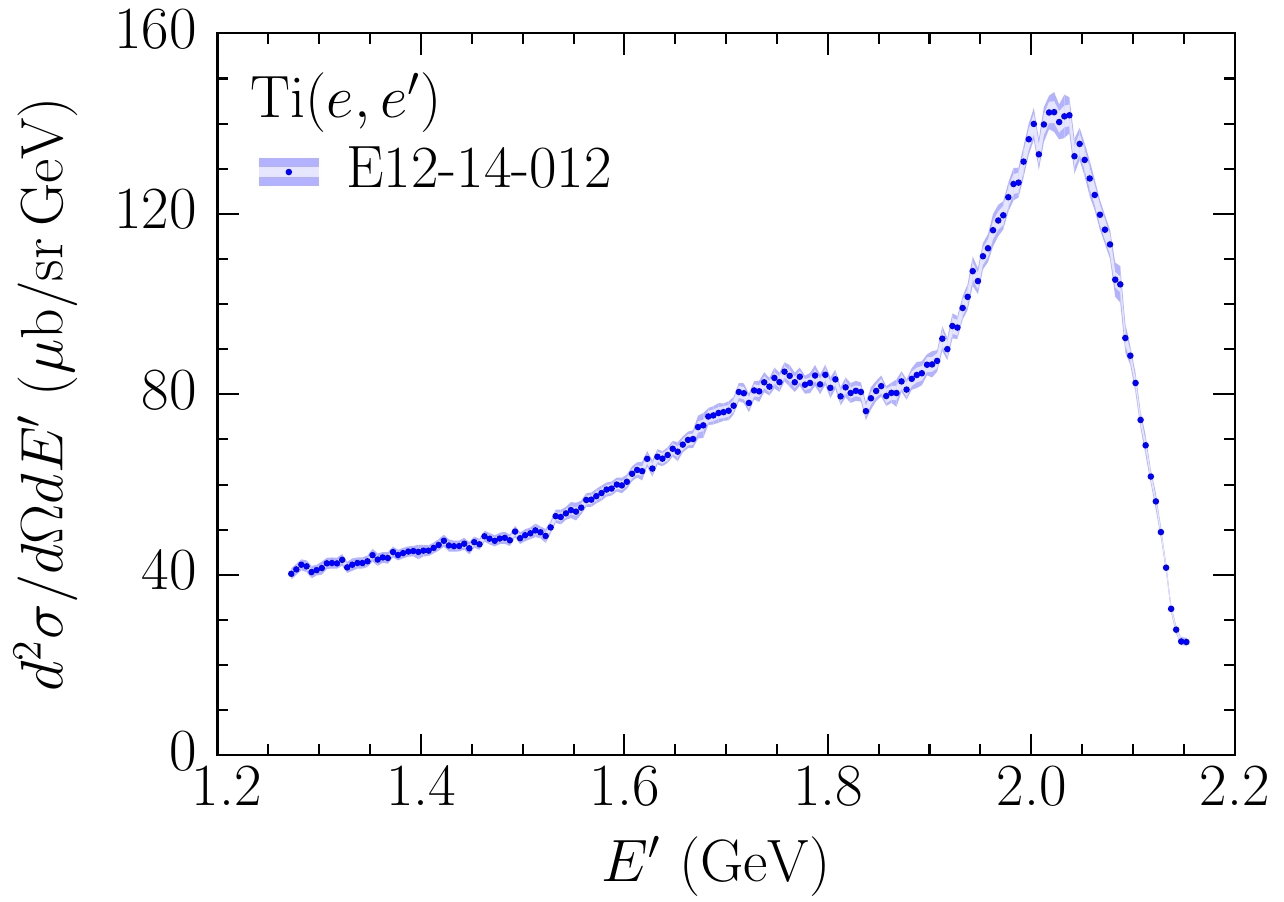}
\caption{(color online). Double differential cross section for the Ti$(e,e^\prime )$ process measured at beam energy of 2.222~GeV and fixed scattering angle of 15.541~deg. The inner and outer uncertainty bars correspond to statistical and total uncertainties, respectively. The maximum uncertainties in the full kinematical range are provided.}
\label{xsec_Ti}
\end{figure}
\par Figure~\ref{xsec_Ti} presents the inclusive electron-titanium cross section, measured at the same kinematics as for carbon and with an error up to
$\sim$2.75\%, sum in quadrature of statistical (1.65\%) and systematic (2.2\%) uncertainties. In the absence of any previous electron-scattering studies carried out using a titanium target, we determined the Ti$(e,e^\prime )$ cross sections using:
\begin{equation}
\left(\frac{d^2\sigma^\text{Born}}{d\Omega dE'}\right)_\text{Ti}^i =
\left(\frac{d^2\sigma^\text{Born}}{d\Omega dE'}\right)_\text{C}^i
\times \frac{\text{Yield}_\text{Ti}^i }{\text{Yield}_\text{C}^i }
\end{equation}
where ${\text{Yield}}^i_{\text{C/Ti}}$ denotes the luminosity normalized yield respectively for C and Ti. By normalizing the yield ratio to published radiatively unfolded carbon cross sections $d\sigma_C^\text{Born}$, we are implicitly unfolding bremsstrahlung from the quoted Ti cross sections.
\begin{figure}
\centering
\includegraphics[width=0.8\columnwidth]{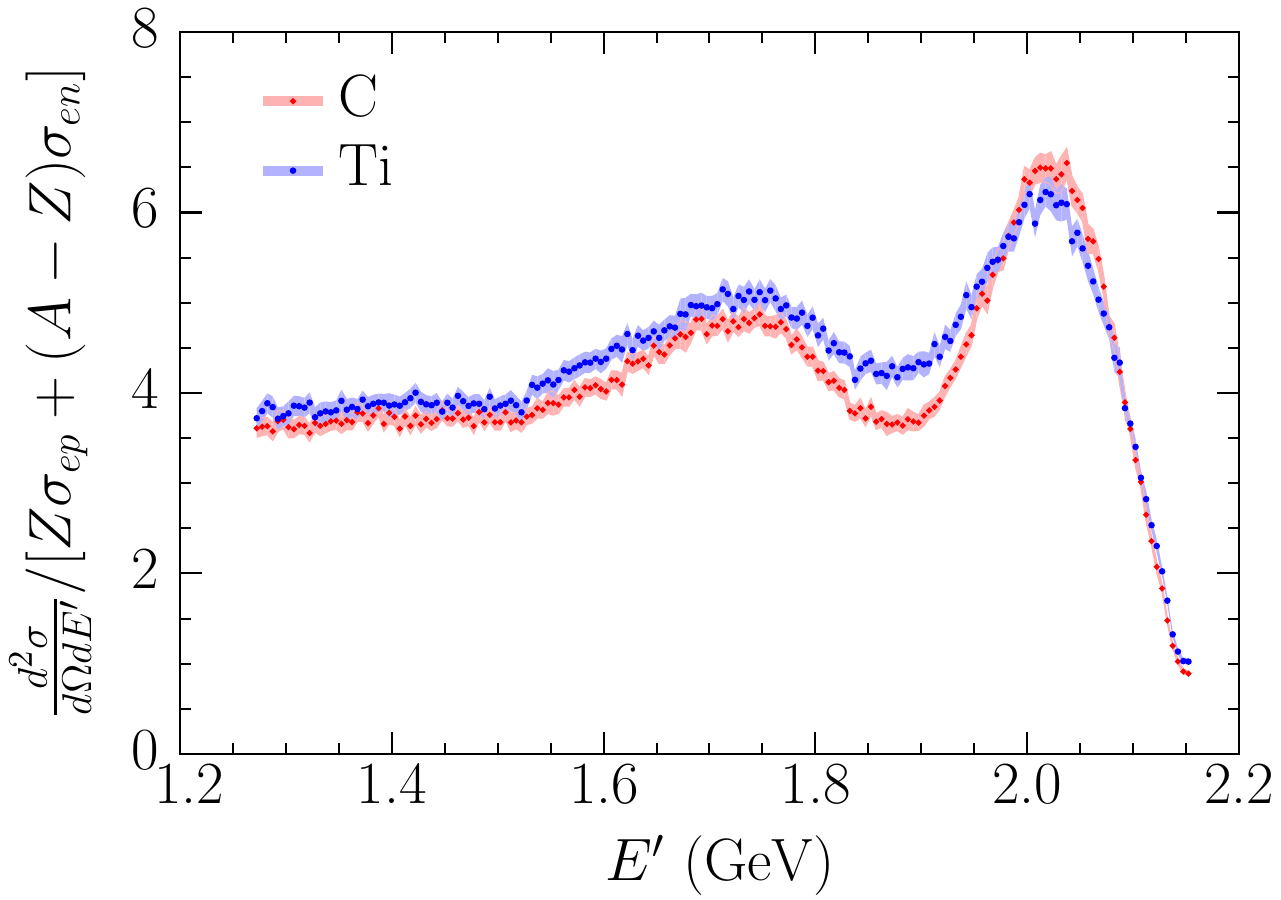}
\caption{(color online). Ratios defined by Eq.\eqref{ratio}, computed using the measured carbon and titanium cross sections. }
\label{xsec_C_Ti}
\end{figure}
In this approach, most of the systematic uncertainties are fully correlated between C and Ti, due to the fact that the data was collected in the same kinematical setup and analyzed using the same cuts of the carbon data. Uncertainties due to radiative corrections, target thickness and density were evaluated independently for Ti, and added in quadrature to the uncertainties from C.
Note that this is the first electron-titanium scattering data collected at the kinematics relevant for neutrino experiments. Therefore, the model of Refs.~\cite{Benhar:2005,Ankowski:2015,Rocco:2016,gosix}, requiring as an input the target spectral function, could not be used to obtain theoretical results comparable to the data of Fig.~\ref{xsec_Ti}. 
\par
\begin{figure}[h!]
\centering
\includegraphics[width=0.8\columnwidth]{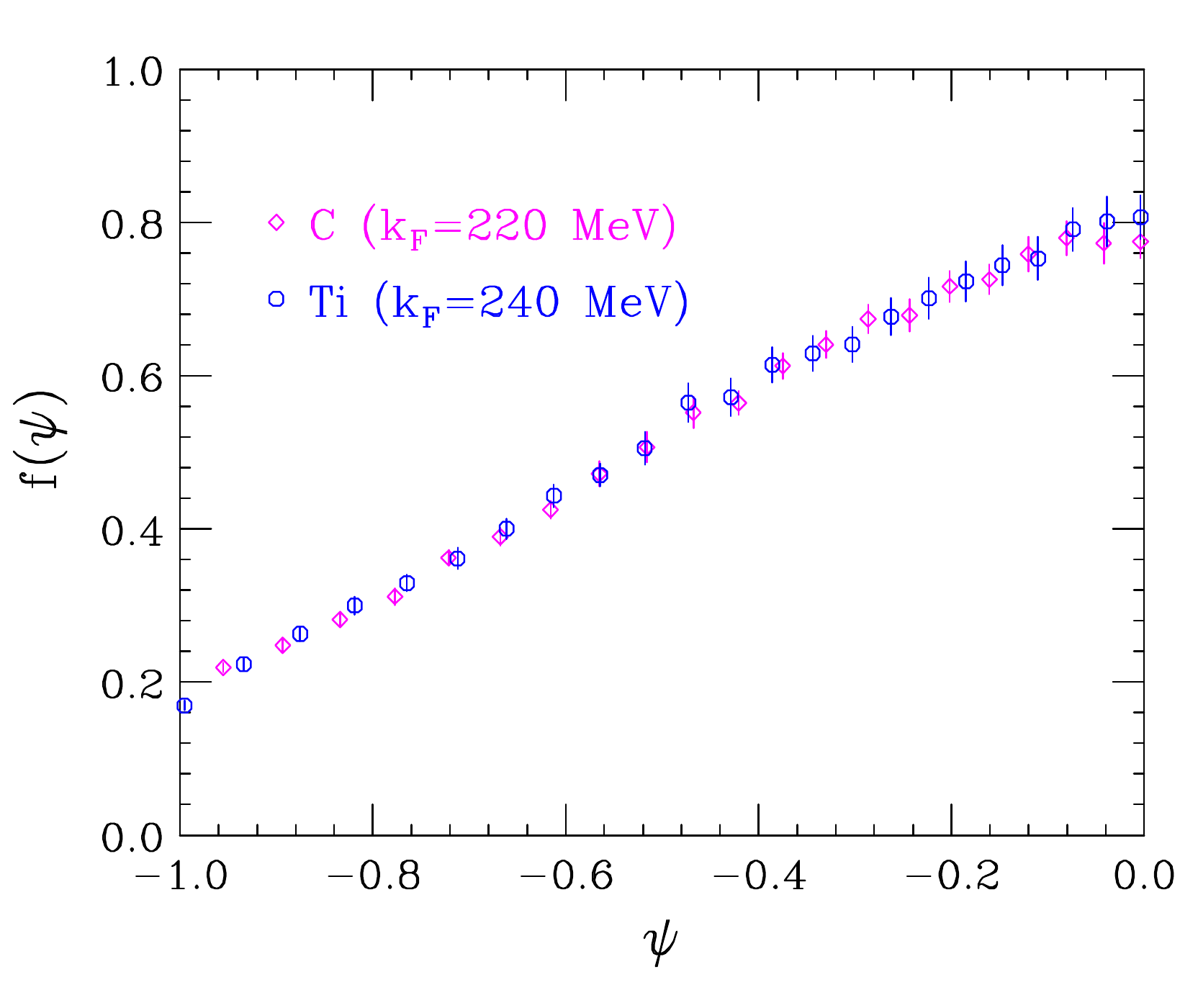}
\caption{(color online). Comparison between the scaling function of $2^\text{nd}$ kind, $f(\psi)$, obtained from the E12-14-012 data on C and Ti, represented by diamonds and 
circles, respsctively. The Fermi momentum of carbon has been fixed to the value obtained by Moniz {\it et al.}~\cite{moniz}. The data analysis for Ti sets the Ti Fermi momentum at $\sim$240~MeV. \\}
\label{fig:superscaling}
\end{figure}

 Figure~\ref{xsec_C_Ti} shows the ratio
\begin{align}
( d^2\sigma/d\Omega d E^\prime ) / [Z\sigma_{ep} + (A-Z)\sigma_{en}] \ ,
\label{ratio}
\end{align}
for carbon and titanium,. Here $\sigma_{ep}$ and $\sigma_{en}$ denote the elastic electron-proton and electron-neutron cross sections stripped of the 
energy-conserving delta function. The difference between the results obtained using the measured carbon and titanium cross sections
reflect different nuclear effects, that can be conveniently parametrized in terms of a nuclear Fermi momentum exploiting the concept of scaling of
second kind, or superscaling~\cite{superscaling}. The superscaling analysis of our data, illustrated in Fig.~\ref{fig:superscaling}, suggests that the 
Fermi momentum in titanium is $\sim$240~MeV, to be compared to 220~MeV in carbon~\cite{moniz}.
%
%
\par In this Letter, we have reported the first results of JLab experiment E12-14-012, consisting of  the {Ti}$(e,e^\prime )$ and {C}$(e,e^\prime )$ cross sections at beam energy $E=$ 2.222 GeV and scattering angle $\theta=$15.541~deg. The quality of the CEBAF electron beam and the excellent performances of the high resolution spectrometer and detector packages available in Hall A allowed for a quick and smooth data taking, and an accurate determination of the cross sections over the broad range of energy transfer in which quasielastic scattering\textemdash induced by both one- and two-nucleon currents\textemdash and resonance production are the main contributions to the inclusive cross sections.
\par Our titanium measurement, providing first electron-titanium scattering data at the kinematics relevant for neutrino experiments, will be of great value for the development of realistic models of the electroweak response of neutron-rich nuclei, which will be indispensable for the analysis of the next generation of neutrino oscillation studies employing argon detectors such as DUNE. Our carbon measurements provide a high precision data that can be utilized in the neutrino experiments that use carbon targets such as T2K, NOvA and MINERvA.
Comparison between the results of theoretical calculations and carbon data confirms that the approach based on the spectral function formalism, supplemented by the inclusion of MEC and FSI contributions,  provides a consistent framework, capable of providing a {\em parameter free} description of electron-nucleus scattering in the kinematical regime in which the IA is expected to be applicable. 
%
%
%
\par We acknowledge the outstanding support from the Jefferson Lab Hall A technical staff, target group and Accelerator Division. The authors are indebted to N.~Rocco for carrying out the calculation of the MEC contribution to the electron-carbon cross section shown in Fig.~\ref{xsec_C}.  This experiment was made possible by Virginia Tech and the National Science Foundation under CAREER grant No. PHY$-$1352106. This work was also supported by the DOE Office of Science, Office of Nuclear Physics, contract DE-AC05-06OR23177, under which Jefferson Science Associates, LLC operates JLab, DOE contracts DE-FG02-96ER40950 and DOE contracts DE-AC02-76SF00515.
%

\end{document}